\documentclass{article}
		\usepackage[english]{babel}
    \usepackage{graphicx}
    \usepackage{subfigure}
    \usepackage{amsfonts}
    \usepackage{amsmath,amssymb}             

		\newcommand{\Det}{{\,\rm Det}}
		\newcommand{\Tr}{{\,\rm Tr}\:}
		
		\newcommand{\eee}[1]{{{\rm e}^{#1}}}
		\newcommand{\D}{{{\hbox{d}}}}
		\newcommand{\mathd}{\mathrm{d}}
		\newcommand{\tmop}[1]{\operatorname{#1}}
		\newcommand{\ham}{{\mathcal H}}

\begin{document}
\hspace*{\stretch{1}}UB-ECM-PF-05/25\\
\hspace*{\stretch{1}}October 2005

\begin{center}
{\Large\bf On gonihedric loops and quantum gravity}

\vspace{4mm}

D.~Espriu\footnote{E-mail: espriu@ecm.ub.es}, 
A.~Prats\footnote{E-mail: prats@ecm.ub.es, address 
after 1st October 2005: prats@lpthe.jussieu.fr}\\
Departament d'Estructura i Constituents de la Materia,\\
Universitat de Barcelona,\\ Diagonal 647, 08028 Barcelona, Spain\\
\end{center}

\begin{abstract}
We present an analysis of the gonihedric loop model, a reformulation
of the two dimensional gonihedric spin model, using two different
techniques. First, the usual regular lattice statistical physics
problem is mapped onto a height model and studied analytically. 
Second, the gravitational version of this loop
model is studied via matrix models techniques. Both methods lead
to the conclusion that the model has $c_{matter}=0$ for all values
of the parameters of the model. In this way it is possible
to understand the absence of a continuous 
transition. 
\end{abstract}

\section{Introduction}

The gonihedric string model (a type of string model with vanishing 
tension at tree 
level) was introduced  and elaborated further in \cite{{Amba-sukiasian}}. 
The proponents argued its potential relevance
to describe the QCD string and is, in any case, an interesting model 
of random surfaces.
The so-called gonihedric
spin model provides a dual version of the gonihedric string embedded in
a (hyper)cubic
discrete space\cite{savvidy,savvidy-bis}.
This spin model describes the gonihedric string as the interface 
between positive and negative spins regions.

The thermodynamical properties of this spin model have been already considered before.
The existence of a second order phase transition separating 
the 'ordered' and 'disordered'\footnote{The quotation marks are 
appropriate here as this model has a exponentially large degeneration and 
so what order and disorder means is not completely apparent.
See \cite{sav}} phases at $T=T_c$ would make 
it possible to
describe the continuum properties of the discrete gonihedric string.
Several papers  studied the phase diagram of this model\cite{baig,esbaijo}, 
but the basic version of the gonihedric spin model exhibited a first
order phase transition rather than a continuous one. 
This led the inclusion of a new
parameter $\kappa$ \cite{savvidy-bis} that regulates the self-avoidance of
the `string'. For $\kappa=0$ the original non-self-avoiding string
theory is recovered, while for increasing $\kappa$ the string becomes
more and more self-avoiding. Of course a non-zero value for $\kappa$ 
is probably of physical interest anyway.
With the addition of this new parameter the phase diagram 
changed considerably
exhibiting second order phase transition for $\kappa$ greater than
some $\kappa_c$.

At some point the spin model gained interest by itself. 
Extremely slow approximation to the equilibrium in a region 
below $T_c$ strongly 
suggested the presence of glassiness in the dynamics
of this three dimensional  model.
The model was studied in \cite{Dim-es-ja-pra}. It revealed indeed the 
presence of
glassiness, when $\kappa=0$, in its dynamics for $T<T_g$, 
where $T_g$ ($T_g<T_c$) 
is the glassy temperature. $T_g$ was determined with good precision 
in the model. 
For $\kappa\neq 0$ it
was seen that the first order transition present at $\kappa=0$ persists
up to some $\kappa_c$. From this point on the transition is of second
order. Also $T_g$ tends to $T_c$ for $\kappa\not=0$ up to a point where
it becomes
impossible to tell them apart; for big enough $\kappa$ the glassy phase
extends all the way up to $T_c$.
The dynamics of a family of very similar models (but not 
including the gonihedric model for any value of $\kappa$) had 
already been studied in
\cite{shore}. 

The very geometrical origin of this spin model fine tunes the coupling 
constants (only nearest neighbors, next to nearest neighbors, and
plaquette-like operators appear) to  very concrete values. With
the addition of the self-avoidance $\kappa$ parameter, a one parameter family
of spin models survives. 
In particular the energy of a spin configuration can be seen to be 
\begin{equation}\label{energy}
E=n_2+\kappa n_4
\end{equation}
where $n_2$ is the length of corners of the inter-phase separating plus
and minus spins, and $n_4$ is the length of the self crossings of this
inter-phase, all in the lattice units. This is in fact the energy of
the gonihedric string, where the flat regions of the string does not
contribute to it at all.

These fine tuned coupling constants are the reason for some crucial
features. The Hamiltonian exhibits a large amount of symmetry; one can
for instance flip at once any whole plane of spins (in three dimensions) 
in any direction
without changing the energy of the configuration (some of this
symmetry is lost when $\kappa\not=0$). 
This feature is thought to
be relevant in the appearance of the glassiness \cite{Dim-es-ja-pra}. The 
appearance of energy barriers along the evolution of the spin
system is another determinant characteristic of this system which is
certainly responsible for very slow dynamics; some metastable states are formed
 where the system gets trapped for a long time. 

In \cite{esp-pra} the model was reduced to two dimensions and the dynamics 
in the two dimensional model studied. Eq. (\ref{energy}) remains valid and
now $n_2$
corresponds to the number of corners and $n_4$ the number of crossings
of the boundary line separating plus and minus domains. This two dimensional
model can actually be represented by a model of loops. We shall term this
the gonihedric loop model.

Although the
two dimensional model exhibits large degeneracy and trapped, long lived, 
metastable states, the dependence of the barriers on the linear scale
of the system is different  and no glassiness was found in this case. 
Only a very
slow, power-like, but certainly faster than logarithmic, evolution to 
equilibrium was
found.  An analytical estimation of the power-like behavior was
computed and found to be in reasonable accordance with numerical simulations.

As for the thermodynamical properties,  
the two dimensional model seems to have
no criticality at all. It is in fact solvable and trivial for $\kappa=0$ (it
actually corresponds to one of Baxter's six vertex models\cite{Baxter}) while
there is no analytical solution  for other
$\kappa$ values. 

In this work we plan to study with different analytical techniques the
gonihedric loop model, both on its own and  coupled to gravity. In this 
latter version the underlying lattice is dynamical. 

In section \ref{loopdef} we briefly recall the basic definitions of the 
gonihedric loop model. In section \ref{loopssect}, we study the 
gonihedric loop model in two dimensions in the framework set 
by Kondev {\it et al.} in \cite{kondev} and by Di Francesco and Guitter 
in a series of papers (see \cite{DiFandGuit} for a review 
and references therein). 
These techniques are developed 
to be used in loop models with no temperature or energy parameter; they
correspond to a combinatorial problem. With some heuristic arguments 
we will connect their techniques and results with our model. 

In section \ref{GonandGRav} we couple the gonihedric loop model 
to gravity. This can be performed using Random Matrix Model techniques, 
more precisely we will write it in terms of a Hermitian two-matrix model. 
This new model is already interesting by itself as it has not been solved 
exactly before 
(although some very similar models have been, see {\cite{Krietal}). 
The peculiar interactions between the two matrices with independent 
coupling constants makes it very hard to solve. Anyway this will provide us
an analytical handle on the gonihedric loop model on randomly generated 
lattices. In the limit $\kappa\to\infty$ one of the matrices in the 
matrix model reduces to a Gaussian variable that can be exactly 
integrated out. Unfortunately this does not make the model directly
solvable. To test then the heuristic arguments we have alluded to before we
study the renormalization group flow of the coupling constants using
the methods in \cite{brezinjus}. As a byproduct we find again confirmation 
of the non-existence of a continuous 
phase transition in the gonihedric spin model 
through the KPZ relations \cite{KPZ}. 

\section{The gonihedric loop model.}\label{loopdef}

First of all we shall present how to construct the loop configurations 
from the gonihedric spin configurations and how to identify the 
points where energy is located.

The gonihedric spin model Hamiltonian has the following form in two dimensions,
\begin{equation} 
\ham_{gonih}^{\textit{\tiny 2D}}=
-\kappa\sum_{<i,j>}\sigma_i\sigma_j 
+\frac{\kappa}{2}\sum_{\ll i,j\gg}\sigma_i\sigma_j
-\frac{1-\kappa}{2}\sum_{[i,j,k,l]}\sigma_i\sigma_j\sigma_k\sigma_l.
\label{Hamiltonian}
\end{equation}
When computing the energy we realize that there are only three possible 
levels of energy for a plaquette. In figure \ref{plaquetteterms} we can 
see all possible configurations up to rotation and $Z_2$ symmetries. 
As we can see in the figure we can associate to each spin configuration 
the inter-phase configuration just drawing a line (dashed line) that 
separates plus and minus spins. This inter-phase will be the gonihedric loop 
and the energy will be accumulated in configurations where the loop  
bends or crosses with itself.

\begin{figure}[!ht]
	\begin{center}
		\subfigure[]{\includegraphics[width=0.20\textwidth]
{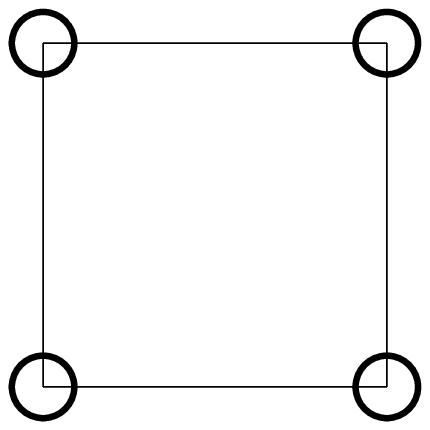}\label{fig1a}}
		\subfigure[]{\includegraphics[width=0.20\textwidth]
{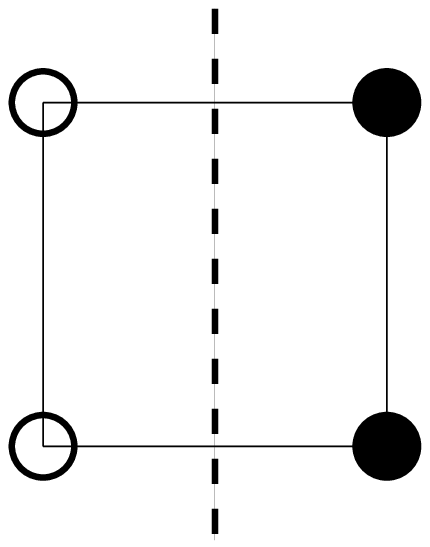}\label{fig1b}}
		\subfigure[]{\includegraphics[width=0.20\textwidth]
{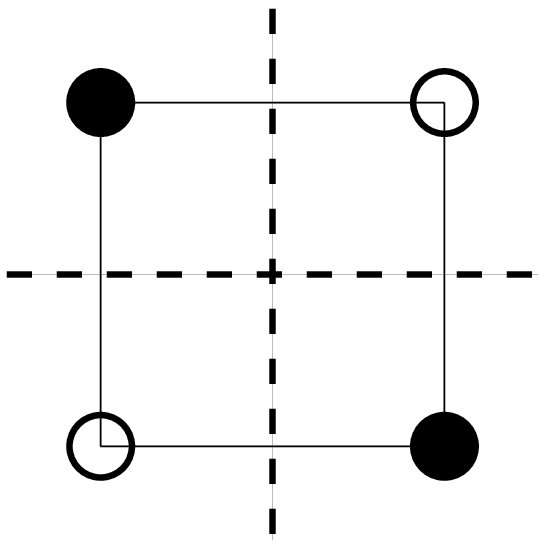}\label{fig1c}}
		\subfigure[]{\includegraphics[width=0.20\textwidth]
{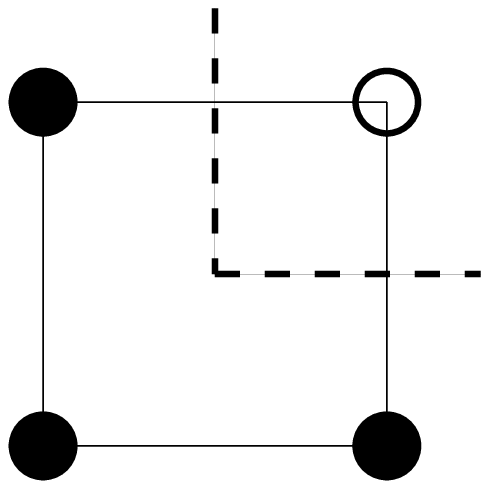}\label{fig1d}}
	\end{center}
	\label{plaquetteterms}
	\caption{All kind of plaquette terms up to symmetries and 
its corresponding loop segment. Figure \ref{fig1a} and \ref{fig1b} 
contributes with some $e_0$ energy while figure \ref{fig1c} 
contributes with $4\kappa$ and figure \ref{fig1d} with $1$ to the energy.}
\end{figure}

Using the Hamiltonian in Eq. (\ref{Hamiltonian}) we compute the
 energy costs to generate bendings and crossings. 
While figure \ref{fig1a} and
\ref{fig1b} cost some bulk energy $e_0$, figure
\ref{fig1c} costs $e_0+4\kappa$ and
\ref{fig1d} costs $e_0+1$ in energy units. That means that all the 
energy will 
accumulate in regions where the loop is bended or it crosses with
itself. Then, apart for a configuration independent term proportional
to the volume, we get an energy per configuration given 
by equation (\ref{energy}).

The Hamiltonian (\ref{Hamiltonian}) thus describes a theory of loops 
defined on a two-dimensional lattice, modulo the usual $Z_2$ degeneracy of 
Ising-like models. The weight of each loop configuration is given by
formula (\ref{energy}).
This is the Gonihedric loop model.
 
Particularly for $\kappa\neq 0$ the model looks like an ordinary Ising model
with some additional interactions. This similarity is in fact way too naive.
In fact the critical behavior as studied
in the references already mentioned is quite different. In this work we shall
understand why.

In fact, when placed on a square regular lattice, 
the above statistical model is trivial for
$\kappa=0$ (only plaquette interactions survive in this case, and we obtain a
model quite different from Ising). For $\kappa\not=0$ there is no known exact
solution of the model.
When we place the above Hamiltonian on
a randomly generated lattice, it will correspond to a model of two
dimensional gravity possibly with some matter contents. The
former case is studied in the next section while the latter is
studied in section \ref{GonandGRav}.

\section{Relation to height models.}\label{loopssect}

In a series of works (see \cite{DiFandGuit} for a review) several aspects of 
loop models in connection with folding and meander problems have been
studied. There is a mapping from loop models onto the so called height models. 
These height models are going to be interesting to us because they 
provide a systematic way to conjecture 
the actual value of the central charge $c$. 
These results are actually based on a previous work by 
Kondev {\it et. al.}\cite{kondev} where the continuous effective Coulomb gas 
formulation is studied.

Let us describe the height models for different classes of loops and recall
some of the results and formalism by Kondev {\it et. al.}. 
We assign to each loop a weight $n=2cos(\pi e)$. The $e=0$ case ($n=2$) 
corresponds 
to the oriented loop model\footnote{A weight $n=2$ may be understood as 
two possible equally probable states for each loop, this could be encoded 
in their orientation.}. The case $e=\frac{1}{3}$, corresponding to a 
weight $n=1$, corresponds to non-oriented loops. 
Let us work first with this $e=0$ case, {\it i.e.} oriented loops.

\subsection{Fully and dense packed loop models on a 
trivalent regular lattice}

Let us begin by considering the so-called fully packed loop model 
on a regular trivalent lattice. This section is based on \cite{DiFandGuit}.
Fully packed means that each site is visited once and only once by a loop. 
We will call $c_{FPL}$ the central charge of this model.
An example of such a loop configuration is drawn in figure \ref{loopconffig}.
\begin{figure}[!ht]
\begin{center}
\includegraphics[scale=0.5]{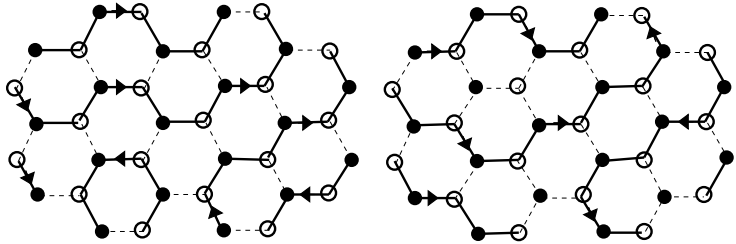} 
\end{center}
\caption{An example of an oriented loop model configuration on a 
bicolorable regular trivalent lattice. The left one is for the fully-packed 
model while the right one is for the dense-packed model. The corresponding 
central charges are $c_{FPL}$ and $c_{DPL}$}
\label{loopconffig}
\end{figure}
The bicolorability of nodes of the lattice enables us to attach 
labels on the loop unit segments. This is equivalent to saying 
that every loop has an even number of unit segments ---then they can 
be labeled using two labels alternatively. We call $B$ the segment 
joining a black node to a white node following the loop orientation, 
and $C$ the segment joining a white node to a black one following the 
loop orientation. We will give a label also to the unvisited edges, 
let us say $A$. In figure \ref{loopconffig} this label association is 
represented. Once we have defined the different type of unit segments 
in our theory we will define the associated height theory.

Define a height $X$ variable in the center of every face. This height 
variable $X$ will differ from one face to another, it actually 
increases by a defined quantity each time we cross a link. This 
quantity is defined following the rules shown in figure \ref{rulesheightfig}.
\begin{figure}[!ht]
\begin{center}
\includegraphics[scale=0.5]{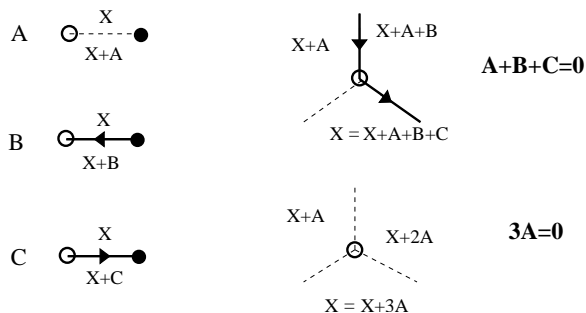}
\end{center}
\caption{Rules defining the increments of the height variable when 
crossing an edge (left) and condition for the height increment unit 
vectors $A$, $B$ and $C$ when going around a vertex (right). 
The bottom one is present only in the dense-packed loop model.}
\label{rulesheightfig}
\end{figure}
Now we just need to ensure that the height variable is globally defined, 
that means that the total gained height in any closed path is zero. 
This amounts in this case to the condition $A+B+C=0$, {\it i.e.} two 
real degrees of freedom remain and so we can describe 
the model with a two-dimensional local field. This implies in the 
continuum limit a model with two scalar fields 
(the two components of the height field) and so the 
central charge for the fully packed loop model $c_{FPL}=2$.

A dense packed loop model is one where the condition of each site being 
visited by a loop is relaxed.
In the case of dense packed loop model we need to consider 
a new kind of vertex. This new vertex is just the non-visited vertex, 
the vacancy or impurity vertex. The condition we find after walking 
around a non-visited vertex is $A=0$, thus $B=-C$, and so the 
associated height model reduces to a one-dimensional height variable.
In the continuum limit the associated continuum height model 
contains only one scalar height field and so in the dense case the 
central charge for the dense-packed loop model is $c_{DPL}=c_{FPL}-1=1$. 
Notice that the $B=-C$ condition erases all information about the 
bicolorability of the lattice. We can see in figure \ref{rulesheightfig} 
that when $B=-C$ it does not matter what color the nodes are.

This was treated in general in the works of Kondev, de Gier and 
Nienhius \cite{kondev}, and used by Di Francesco and 
Guitter \cite{DiFandGuit}. 
In the general picture a weight $n=2\cos(\pi e)$ per loop ($e\in [0,1]$) is 
introduced and the the model is studied following an effective field theory 
description. This analysis gives 
\begin{equation}\label{conformalc}
c_{DPL}=c_{FPL}-1=1-6\frac{e^2}{1-e}.
\end{equation}

For the non-oriented case, equivalent as we have already said 
to $e=\frac{1}{3}$ in the general description, we can still analyze 
the height components in the same form. 
We have no orientation so we cannot use the bicolorability 
information to define in a meaningful way the labels on the segments. 
Then we get only two kind of height increments, say $A$ and $B$. 
\begin{figure}[!ht]
\begin{center}
\includegraphics[scale=0.5]{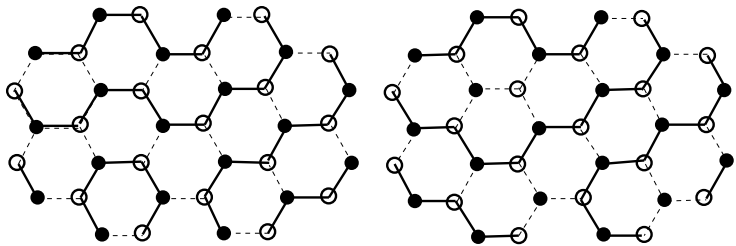} 
\end{center}
\caption{An example of a non-oriented loop model configuration on a 
bicolorable regular trivalent lattice. Fully-packed loop model 
on the left and dense-packed loop model on the right.}
\label{loopconffig2}
\end{figure}
In figure \ref{loopconffig2} we can see that there is no way to 
distinguish different kind of segments on the loop. 
The bicolorability property combines with the orientable nature 
of the loops to define the $B$ and $C$ increment vectors. 
The loss of any of the two ingredient forces a constraint between $B$ and $C$. 
The rules for the height increments are shown in figure \ref{loopconffig3}. 

In the fully packed non-oriented loop model the only vertex 
we have is again a visited vertex with two loop segments and an 
empty edge. The condition ensuring a meaningful height variable 
is thus $A+2B=0$. Following again the same steps as before we end 
up with a one dimensional height variable, meaning that $c_{FPL}=1$ 
in the non-oriented loop model on a trivalent regular lattice.

The dense packed non-oriented loop model is again straightforward. 
The extra condition for the new unvisited vertex is $3A=0$. 
This constraints $A$ and $B$ to zero so the height variable remains 
constant in the whole lattice. This means no scalar field in the 
continuum theory, and consequently $c_{DPL}=c_{FPL}-1=0$. 
Note that this coincides with the values we would obtain 
from equation (\ref{conformalc}) when applying $e=\frac{1}{3}$.
\begin{figure}[!ht]
\begin{center}
\includegraphics[scale=0.5]{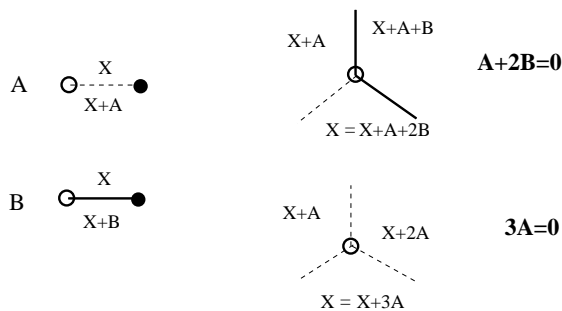} 
\end{center}
\caption{Left: In this case the height increment rules involve 
only two increment vectors, $A$ and $B$. Right: The conditions on this vectors}
\label{loopconffig3}
\end{figure}

These two examples help us to understand how the height models 
are constructed, and how to extract the information we are looking for.

\subsection{Non-oriented dense  loops 
on tetravalent regular lattice.}
Taking  the gonihedric spin model as our starting point we can arrive
 at a loop model 
similar to the ones we have just described. One difference 
is that our spin model is placed in a square lattice and 
so are the loops that live on the dual lattice. 
As already explained above these loops will add energy to the system 
due to  bending ($\Delta E= 1$) and crossing ($\Delta E=4\kappa$). 

Let us consider first the case where $\kappa=\infty$ 
(infinite self-avoidance). In this case there is no crossing of the loops, 
so if we forget for a moment about the temperature we have a model of 
dense loops with no self crossings. They are of course not oriented 
loops since there is no way the spin configuration can induce an independent
orientation to each loop.
The case where $\kappa<\infty$ can be recovered from the $\kappa=\infty$ case 
just by relaxing the no-crossings condition. In the opposite limit ($\kappa=0$)
 crossing configurations occur freely. We will argue later how 
to extract conclusions for the whole range of $\kappa$.

In figure \ref{loopconfgonihedric} we see loop configurations without and
with intersections. 
The rules needed to label all (occupied or not) edges, and the 
possible vertex present in our model are shown in 
figure \ref{loopconfgonihedric_2}. Since our loops are not oriented 
we cannot define in a consistent way the alternated labeling of the 
segments of the loop. This is why all segments in the loop have to have 
the same label $B$. In the same way, considering the `dual loop' 
constructed from the unvisited edges, we see easily that all 
unvisited vertexes have the same label $A\not=B$.
\begin{figure}[!ht]
\begin{center}
\includegraphics[scale=0.5]{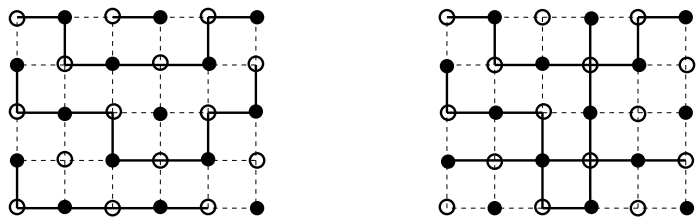} 
\end{center}
\caption{An example of a dense-packed non-oriented loop model configuration 
on a bicolorable regular tetravalent lattice. On left hand side 
there is a non-self-intersecting loop model while right 
hand side the example shows a configuration with self-crossing loops.}
\label{loopconfgonihedric}
\end{figure}
Now, looking at the total height difference when going around one 
site in our lattice, we can extract some consistency conditions 
for our height unit increments $A$ and $B$. 
These local consistency conditions are 
depicted in figure \ref{loopconfgonihedric_2}.
It is easy to see then that the equations required from consistency of the
height variables
\begin{eqnarray}
2A+2B&=&0\\
4A&=&0\\
4B&=&0\label{crosseq}
\end{eqnarray}
\begin{figure}[!ht]
\begin{center}
\includegraphics[scale=0.5]{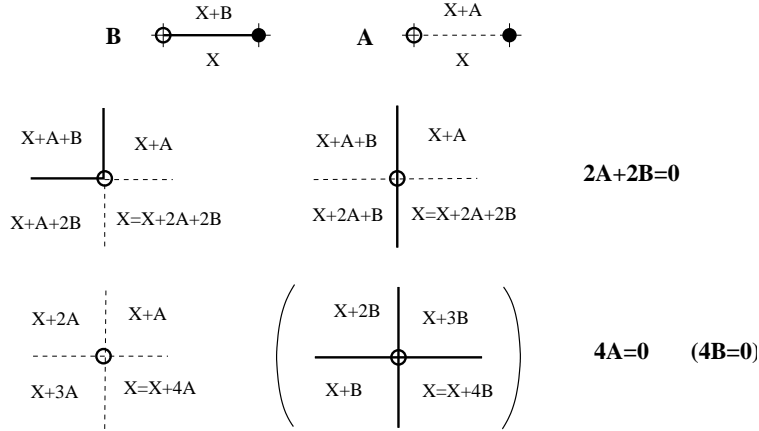} 
\end{center}
\caption{Rules defining the height increments 
from the loop configuration and consistency equations from 
the different kind of vertex. In parenthesis the vertex 
appearing with the self-intersections.}
\label{loopconfgonihedric_2}
\end{figure}
constrain our height variable so much that it is in fact 
constant all over the lattice. 

We should remember that there is no temperature or Hamiltonian 
in these loops models from which we extract the height model. 
The connection with our model appears when we consider the relevance 
of the bending vertex. We obtain the same equations for $A$ and $B$ 
either allowing or not the bending of the loops. 
This corresponds to studying our loops inherited from 
the gonihedric spin model at $T=0$ or $T=\infty$, 
respectively \footnote{Note that in the loop/height model, a bending 
vertex has the same weight as the straight loop vertex and the bulk. 
This corresponds to infinite temperature. When we do not allow bending 
to appear, 
it is equivalent to freezing the gonihedric loops at $T=0$.}. 
We should then come to the conclusion that there is no continuous
transition at finite temperature since both extrema 
share the same critical behavior\cite{c-theorem}. 
The same reasoning with the self-crossing vertex being allowed 
or not brings us to the same conclusion for every value of the 
coupling constant $\kappa$ in the gonihedric 
Hamiltonian\footnote{Remember that $\kappa=0$ is in fact 
trivially solved \cite{Baxter} and does not possess  
any thermodynamical transition.}.
Thus we end up again in a $c=0$ theory in the whole $\kappa$ space. 
This means no criticality, in perfect agreement with numerical 
simulations\cite{esp-pra}.

\section{The two-dimensional gonihedric spin model 
coupled to gravity}\label{GonandGRav}

We are now going to couple the gonihedric loop model to gravity. 
This is interesting by itself and in order to understand the effect of 
placing the gonihedric spin model 
on a 'fluid' surface. Of course the resulting model is nothing but 
the one-dimensional version 
of the gonihedric string in a two-dimensional embedding curved 
dynamical space.\footnote{In higher dimensional cases related studies in the
continuum have been performed by \cite{Fazio:2003sv}.}

The energy of the gonihedric spin model can still be written, on a 
fixed quadrangulation, as $n_2+4\kappa n_4$, where $n_2$ is the 
number of unit 'corners' of the inter-phase between plus and minus 
sign spins, and $n_4$ is the number of self-crossing points of 
the inter-phase. 

In order to couple this model  to gravity
 we need to put our spins in the sites of a 
random lattice built from square pieces\footnote{We chose squares 
as the building blocks of our random lattice since our original model 
is defined in a square regular lattice. It happens to be important in order 
to define what exactly means bending the loop.}. 
We will consider, when coupling the model  to gravity, the 
limiting case where the loops 
\begin{figure}[!ht]
\begin{center}
\includegraphics[scale=0.3]{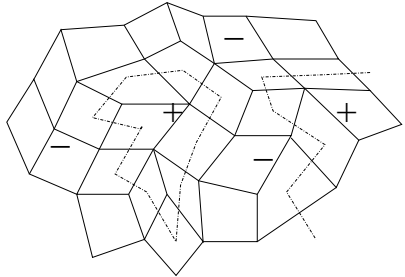}
\end{center}
\caption{Example of a random lattice with a gonihedric spin model on it}
\label{random lattice}
\end{figure}   
never cross themselves.  
This non intersecting limit corresponds to the $\kappa\to\infty$ limit 
in the gonihedric spin action.
In fig.\ref{random lattice} we can see an example of this kind of 
quadrangulations. 

To describe these configurations we shall introduce $N\times N$ Hermitian 
random matrices. Let us see how to do this.
We have first of all plaquettes 
where no loop goes across them, then  
plaquettes crossed by one loop without bending 
through it (a straight loop piece across a unit square of the lattice), 
and finally plaquettes were the loop crossing them bends 'right' 
or 'left'. These three building blocks of the random lattices 
are presented graphically\footnote{To simplify the visual identification 
with the loop model we have not drawn the lines corresponding to the bulk 
propagator, i.e. to the A matrix propagator. The loop 
is generated with the B matrix propagator.} in fig.\ref{intterms} 
with the corresponding term of the associated matrix model 
that will generate them upon integration.
\begin{figure}[!ht]
\begin{center}
\includegraphics[scale=0.3]{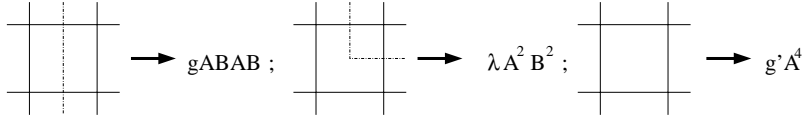}
\end{center}
\caption{Correspondence between the loop pieces and the matrix
  interaction that are going to generate them.}
\label{intterms}
\end{figure}
As we can see we are considering the most general case where all the
couplings are different. In our specific case we will impose the
condition $g'=g$ due to the fact that a straight piece of loop
does not contribute with any amount of energy to the action, so the
coupling needs to be equal to the bulk coupling.
In addition to these 'interaction' terms we need to include
 the kinetic term for the two
matrices, i.e. the quadratic terms $A^2$ and
$B^2$ that generates the propagators. So finally the matrix 
model that will represent our gonihedric 
loop model coupled to gravity is
\begin{equation}
{\textsl{Z}}_N(g,\lambda)=\int^{(N)}\mathrm{d}A\mathrm{d} B 
\exp\left[-N\Tr\left(\frac{1}{2}(A^2
+B^2)+\frac{g}{4} A^4+\frac{g}{2} ABAB+\lambda A^2B^2\right)\right] \nonumber
\end{equation}
The numerical factors preceding each interaction term are in 
fact symmetry factors. To our knowledge this model has not been 
solved exactly. Although somewhat similar matrix models have 
been solved  elsewhere\cite{Krietal}, their
solution cannot be applied directly to our matrix model. 

The coupling constants $g$, $g'$ and $\lambda$ have 
a definite meaning in terms of the statistical model 
we are treating. $g=g'$ are in fact representing the bulk 
energy of the unit square, {\it i.e.} it represents the conjugate 
variable to the area of the lattice, the cosmological constant. 
On the other side the combination $\frac{\lambda}{g}$ is equal 
to the statistical Boltzmann weight of each bended loop piece, 
that contains all the information from the energy of the statistical model.
\begin{equation}
g=g'=\frac{\eee{-\beta E_{bulk}}}{N} \ ,
\qquad \frac{\lambda}{g}=\eee{-\beta E_{bend}}
\end{equation} 

\subsection{Partial integration}

The $B$ matrix integration is in fact Gaussian 
so it can be integrated exactly. This integration was first presented 
in \cite{Polonica}. We first diagonalize the $A$ matrix 
\begin{eqnarray}
{\textsl{Z}}_N(g,\lambda)&=&\int\mathrm{d}A\mathrm{d} B 
\exp\left[-N\Tr\left(\frac{1}{2}(A^2
+B^2)+\frac{g}{4} A^4+\frac{g}{2} ABAB+\lambda A^2B^2\right)\right] \nonumber\\
&=&\int\left(\prod_{i=1}^N\mathrm{d}a_i\D B_{ii}\right)
\left(\prod_{j>i=1}^N\mathrm{d} \mathrm{Re}B_{ij}\mathrm{d} 
\mathrm{Im}B_{ij}\right) 
\prod_{j>i=1}^N(a_j-a_i)^2\eee{-NS(\{a_i\},\{B_{ij}\})}
\end{eqnarray}
where $\{a_i\}$ are the eigenvalues of A. The action $S(\{a_i\},\{B_{ij}\})$ is
\begin{eqnarray}
S(\{a_i\},\{B_{ij}\})&=&\sum_{i=1}^N\left(\frac{1}{2}a_i^2+
\frac{g}{4} a_i^4\right)+\frac{1}{2}\sum_{i=1}^N B_{ii}^2
\left(1+(g+2\lambda)a_i^2\right)\\
&&+\sum_{j>i=1}^N(\mathrm{Re}B_{ij}^2+\mathrm{Im}B_{ij}^2)
\left(ga_ia_j+\lambda(a_i^2+a_j^2)\right)
\end{eqnarray}
Rescaling the $B$ matrix entries correspondingly we transform 
the $B$ integral in a bunch of decoupled Gaussian integrals
\begin{eqnarray}
{\textsl{Z}}_N(g,\lambda)&\sim&\int\prod_{i=1}^N\mathrm{d}a_i
\prod_{j>i=1}^N(a_j-a_i)^2 \prod_{i,j=1}^N(\delta_{ij}+ga_ia_j+
\lambda(a_i^2+a_j^2))^{-1/2}\nonumber\eee{-N\sum_{i=1}^N
\left(\frac{1}{2}a_i^2+\frac{g}{4} a_i^4\right)}\\
&=&\int\D A \left(\Det\left[\mathbb{I}
\otimes\mathbb{I}+gA\otimes A +\lambda(A^2\otimes\mathbb{I}+
\mathbb{I}\otimes A^2)\right]\right)^{-\frac{1}{2}} 
\eee{-N\Tr\left(\frac{1}{2}A^2+\frac{g}{4}A^4\right)}\label{onematrred}
\end{eqnarray}
where $\Det$ means that a determinant of a tensor 
product matrix, {\it i.e.} a $N^2\times N^2$ matrix. 

This is a nice expression of the two matrix model 
as a one matrix model. Although some especial cases of this model 
with specially fine tuned coupling constants have been solved, 
it appears to be very difficult to solve in the general parameter setting 
which is interesting to us. So we have to turn to other techniques.

\section{The renormalization group approach}

The renormalization group approach for random matrices was first 
introduced in \cite{brezinjus} by Br\'ezin and Zinn-Justin and later 
improved by Higuchi, Itoi, Nishigaki and Sakai in \cite{Higu}. 
The improvement in \cite{Higu} provides a more quantitative 
analysis of the critical structure in parameter space. 
It consists basically in using the equations of motion of the matrix 
model (also called loop equations) in order to deal with redundant 
operators in the renormalized action in the large $N$ limit. 
This more complete analysis is much more involved and we have 
not applied it to our model but instead used the simpler 
Br\'ezin--Zinn-Justin approach since it is equally valid to get the 
qualitative flow we are looking for. 

The renormalization procedure in matrix models consists in 
the integration of the last row and column of the matrices, so that 
we relate the partition function $\textsl{Z}^{(N+1)}$ to $\textsl{Z}^{(N)}$. 
One could in principle do so with the one-matrix reduced model 
(\ref{onematrred}), but due to non linearities in $\Tr(A)$ and $\Tr(A^2)$ 
already in the saddle point equation we decided to return to the original 
two-matrix model, where this non-linearity is not present. 
In order to make easier the `last-row--last-column' integration 
we are going to write the matrices $A$ and $B$ as follows
\begin{eqnarray}
A_{N+1}&=&\left(\begin{array}{cc}
	A_N& 0\\
	0&\alpha
\end{array}\right)\\
B_{N+1}&=&\left(\begin{array}{cc}
	B_N& \vec{v}\\
	{\vec{v}\,}^\dagger&\beta
\end{array}\right)\label{reducedform}
\end{eqnarray}
Since the action is Gaussian in the $B$ matrix we know that the integral 
in the $\vec{v}$ vector and that of the $\beta$ variable will be easy 
to perform. Only a saddle point in the $\alpha$ variable will be left 
as in the original approach \cite{brezinjus}.

To perform the renormalization group computations we need 
to free our parameters from the constraint that the original statistical
model imposes on them, {\it i.e.} we should let $g\not=g'$. 
The partition function will satisfy
\begin{equation}
\textsl{Z}_{N+1}(g,g',\lambda)=C(N)^{N^2}\,
\textsl{Z}_{N}(g+\delta g,g'+\delta g',\lambda+\delta\lambda)
\end{equation}
where $\delta g$, $\delta g'$ and $\delta\lambda$ depend on $N$. 
This equation will be well approximated by the linear differential equation 
\begin{equation}
\left[N\frac{\partial}{\partial N}-\beta_g
\frac{\partial}{\partial g}-\beta_{g'}
\frac{\partial}{\partial g'}-\beta_\lambda
\frac{\partial}{\partial \lambda}+\gamma\right]
F(N,g,g',\lambda)=r(N)
\end{equation}
where $r(N)$ is related to $C(N)$ and 
\begin{eqnarray}
F(N,g,g',\lambda)=-\frac{1}{N^2}\log 
\left[\frac{\textsl{Z}_N(g,g',\lambda)}{\textsl{Z}_N(0,0,0)}\right]
\end{eqnarray}
is the free energy of our model.

\subsection{Last-column-last-row integration}

Consider
\begin{eqnarray*}
  \textsl{Z}_{N + 1} ( g, g',\lambda ) & = & \int^{(N + 1)} 
\mathd A \mathd B\,\, \eee{ - (N + 1) \tmop{Tr}_{N + 1}  
[\frac{1}{2}(A^2 + B^2) + \frac{g'}{4} A^4 + \frac{g}{2} A B A B + 
\lambda A^2 B^2 ]}.
\end{eqnarray*}
We rotate both matrices with the same $U\in U(n)$ unitary matrix 
in such a way that the $A$ matrix diagonalize. 
One unitary integral decouples and contributes with a 
factor $V_{U(N+1)}=\pi^{N(N+1)/2}/\prod_{p=1}^{N+1}p!$. The matrix $A$ is now 
\begin{eqnarray*}
  A_{N+1} & = & \left(\begin{array}{cccc}
    a_1 &  &  & 0\\
    & \ddots &  & \\
    &  & a_N & \\
    0 &  &  & \alpha
  \end{array}\right)
\end{eqnarray*}
After this we rotate back the first $N$ rows and columns so that $A$ 
and $B$ will look like Eq. (\ref{reducedform}).

It is easy to see that all the interaction terms in the action 
can be written as
\begin{eqnarray*}
  \tmop{Tr}_{N + 1} [ A^n ] & = & \tmop{Tr}_N [ A^n ] + \alpha^n\\
  \tmop{Tr}_{N + 1} [ B^2 ] & = & \tmop{Tr}_N [ B^2 ] + \beta^2 + 2
  \vec{v}^{\dag} \vec{v}\\
  \tmop{Tr}_{N + 1} [ A B A B ] & = & \tmop{Tr}_N [
  A B A B ] + 2 \alpha \vec{v}^{\dag} A \vec{v}
  + \alpha^2 \beta^2\\
  \tmop{Tr}_{N + 1} [ A^2 B^2 ] & = & \tmop{Tr}_N [ A^2 B^2 ] + \alpha^2 (
  \vec{v}^{\dag} \vec{v} + \beta^2 ) + \vec{v}^{\dag} A^2 \vec{v}.
\end{eqnarray*}
After substituting this into the partition function 
we  find several terms. 
Those depending on $\vec{v}$, $\beta$, and $\alpha$ will be integrated 
out so that a new effective action for the $N\times N$ two-hermitian 
matrix model will arise. 
After performing the integrations on $\vec{v}$ and $\beta$ we find 
\begin{equation}
  \textsl{Z}_{N + 1} ( g,g', \lambda ) = \frac{V_{U(N + 1)}}{V_{U(N)}} 
C_N \int^N \mathd A \mathd B \,\eee{ -
  (N + 1) \tmop{Tr}_N \left[ \frac{1}{2}(A^2 + B^2) + 
\frac{g'}{4} A^4 + \frac{g}{2} ABAB + \lambda A^2 B^2 \right]} 
  I(g,g^\prime,\lambda), 
\end{equation} 
where
\begin{equation}
C_N=\left(\frac{\pi}{N+1}\right)^N\left(\frac{2\pi}{N}\right)^{\frac{1}{2}},
\end{equation}
and
\begin{equation}
I(g,g^\prime,\lambda)=
\int \mathd \alpha \frac{\det [A-\alpha]^2} 
{\det[1+\lambda\alpha^2+g\alpha A+\lambda A^2]
(1+(2\lambda + g)\alpha^2)^{\frac{1}{2}}} 
\eee{- (N + 1)\left(\frac{\alpha^2}{2} + \frac{g'}{4} \alpha^4\right)}.
\label{integral1} 
\end{equation}

We can already see here that we are not going to receive any direct 
contribution to the renormalization of the $\lambda$ or $g$ constant 
in the action. There is no $B$ matrix dependence in the saddle point integral, 
so we cannot generate the operators $\Tr(A^2B^2)$ or $\Tr(ABAB)$ in our 
renormalization scheme. The only flow of this two parameters comes from the 
renormalization of the ``kinetic'' $\Tr(A^2)$ operator after field rescaling.
The $\alpha$ integral cannot be computed explicitly. 
The saddle point method has been used to solve it in the large $N$ limit. 
The two determinants in the integrand contribute to the saddle point 
while the square root in the denominator does not.

\subsection{Saddle point equation}

\noindent The saddle point equation reads
\begin{eqnarray}
\frac{2}{N}\Tr\left(\frac{1}{\alpha_s-A}\right)=\left(\alpha_s+g'\alpha_s^3
\right)+\frac{1}{N} \Tr\left(\frac{2\lambda\alpha_s+g A} 
{1+\lambda\alpha_s^2+g\alpha_s A+\lambda A^2}\right).
\end{eqnarray}
To solve this saddle point equation we will assume that $\alpha_s$,
the solution of the saddle point equation, has an expansion in 
powers of $\hat a_j=\frac{1}{N}\Tr(A^j)$. Once this saddle point is computed 
we must introduce $\alpha_s$ into the integral, then we will again expand 
everything to find the contribution of this integral to the operator $\hat a_2$
\begin{eqnarray}\label{integralsaddle}
I(g,g^\prime,\lambda)
&=&\frac{\det [A-\alpha_s]^2} {\det[1+\lambda\alpha_s^2+g\alpha_s A+
\lambda A^2]
(1+(2\lambda + g)\alpha_s^2)^{\frac{1}{2}}} 
\exp{- (N + 1)\left(\frac{\alpha_s^2}{2} + \frac{g'}{4} \alpha_s^4\right)}\\
&=&\exp{-(N+1)\left(C_{0}+C_{1}\hat a_1+C_{1,1}\hat a_1^2+C_2 \hat a_2+
\cdots\right)}
\end{eqnarray}
To keep track of the terms we need in the expressions we will perform 
the change of variables $A\to\epsilon A$. At the end of the calculation we 
shall put $\epsilon=1$.
\begin{eqnarray}
\frac{2}{N}\Tr\left(\frac{1}{\alpha_s-\epsilon A}\right)=
\left(\alpha_s+g'\alpha_s^3\right)+\frac{1}{N} \Tr\left(\frac{2\lambda\alpha_s
+g \epsilon A} {1+\lambda\alpha_s^2+g\alpha_s \epsilon A+\lambda 
\epsilon ^2 A^2}\right).
\end{eqnarray}
We are going to expand everything up to second order in $\epsilon$. 
This will be enough to extract the $C_2$ coefficient.
We expand
\begin{equation}
  \alpha_s = \alpha_s^{( 0 )} + \epsilon \tilde\alpha_s^{( 1 )} + 
\epsilon^2 \tilde\alpha_s^{( 2 )}
  +\mathcal{O}( \epsilon^3 ),
\label{expansion}
\end{equation}
where
\begin{equation}
\tilde\alpha_s^{(1)}=\alpha_s^{(1)} \hat a_1,\qquad
\tilde\alpha_s^{(2)}=\alpha_s^{(2)} \hat a_2 + \alpha_s^{(1,1)} {\hat a_1}^2.
\end{equation}
At order $\mathcal{O}( \epsilon^0 )$ we get an equation for $\alpha_s^{(0)}$
\begin{equation}
\alpha_s^{(0)}+g'\alpha_s^{(0)3}-\frac{2}{\alpha_s^{(0)}+
\lambda\alpha_s^{(0)3}}=0\label{6thorder}
\end{equation}
that is a sixth order equation reducible to a third order one. Of 
all solutions we will chose the one that matches the known 
solution for $g\to 0$ and $\lambda\to 0$, 
that is $\alpha_s^{(0)}\to \sqrt{2}$
\footnote{We could also choose $-\sqrt{2}$.} and which is real.
At next order we get a linear equations for $\alpha_s^{(1)}$, and at 
order $\mathcal{O}( \epsilon^2 )$ we get two linear equations.
We have tested the stability of the saddle point solution we have found 
in the vicinity of the point $g=\lambda=0$ using the above expansion. 
We have checked that the solution is stable for small values of 
these coupling constants in the whole $g'$ range.

At this point we can express the saddle point evaluation of the integral 
in terms of these constants and find the renormalization of the 
operator $\Tr(A^2)=N\hat a_2$.
If we call $Z_2$ the coupling relative to that operator (when we begin 
$Z_2=1$) 
the renormalized coupling $Z_2^{ren}$ after one renormalization group step 
from the original model will be
\begin{equation}
Z_2^{ren}=\frac{N+1}{N}\left(1+\frac{1}{N+1}C_2\right),
\end{equation}
where
\begin{equation}
C_2= \frac{2+6\alpha_0^2-(g^2-4\lambda^2)\alpha_0^4}
{2(\alpha_0+\lambda\alpha_0^3)^2}
\end{equation}
and $\alpha_0$ is the solution we took from the sixth order equation Eq.
(\ref{6thorder}).

\subsection{Renormalization-group flow}

We must scale the $A$ matrix in order to keep $Z_2^{ren}=1$ at the 
end of the complete renormalization group process. 
This induces a flow on the couplings $\lambda$ and $g$. 
\begin{eqnarray}
\lambda^{ren}&=&\lambda\left(1-\frac{1}{N}C_2+\mathcal{O}(N^{-2})\right)\\
g^{ren}&=&g\left(1-\frac{1}{N}C_2+\mathcal{O}(N^{-2})\right)
\end{eqnarray}
The behavior under the renormalization-group depends on the sign of $C_2$. 

If we pay attention to the statistical meaning of the couplings in the 
original model, it is
clear that we can keep $g\ll 1$ and both $g$ and  $\lambda/g$ are positive,  
{\it i.e.}  $\lambda,g>0$ is the physical quadrant. 
As in the usual renormalization group approach 
$\frac{\delta (g-g^*)}{g-g^*}<0$ near $g=g^*$ implies an infrared fixed point 
at $g=g^*$. This is the case in the physical quadrant.

In other words, the two constants $\lambda$ and $g$ renormalize 
towards its fixed value $\lambda=g=0$. Thus only the $g'$ coupling constant 
remains. The constant $g'$ accompanies is the operator $\Tr(A^4)$  which 
generates 
only pure gravity critical behavior.  This means $c_{matter}=0$. 
\begin{figure}[!ht]
\begin{center}
\subfigure[]{\includegraphics[scale=0.65]{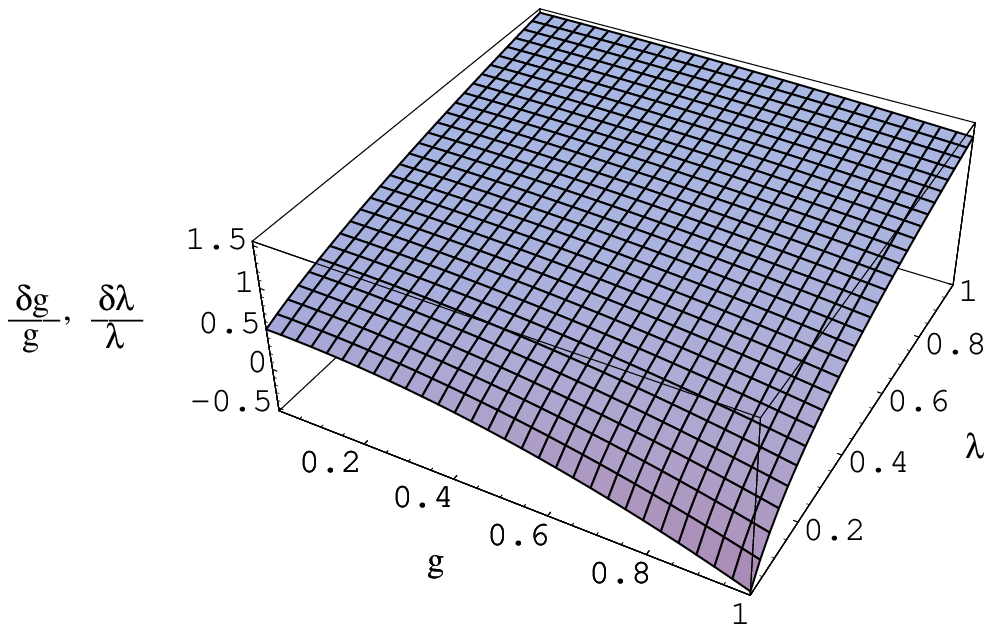}}\hspace{1cm}
\subfigure[]{\includegraphics[scale=0.65]{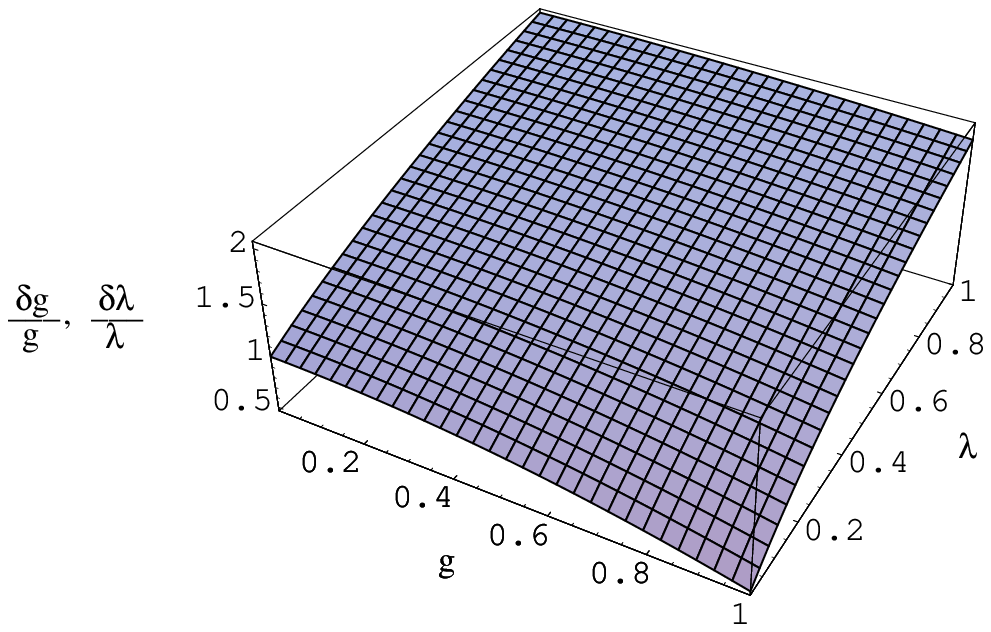}}
\caption{$\frac{\delta g}{g}$ and $\frac{\delta\lambda}{\lambda}$ 
for positives 
values of the coupling constants around the origin (physical quadrant). 
Two different 
values of $g'$ have been plotted to show the positiveness of this 
quantity for all $g'$. (a) $g'\sim 0$. (b) $g'=1$.}
\end{center}
\label{renorm}
\end{figure}

We can also think of this flow in the following way. Regardless of $g$, 
if $\lambda$ is flowing towards zero, we 
can forget about the corresponding 
operator and concentrate on the rest of the action. 
The corresponding model has been solved in \cite{zinnjus}. 
In our special choice of parameters $g'=g$ this brings us to pure 
gravity again.

The irrelevance of the operator $\Tr(A^2B^2)$ could have also been already 
predicted with the heuristic tools presented in the first half part of this 
paper. Using the height model language from \cite{DiFandGuit}, 
we can see that for ordinary gravity\footnote{respect Eulerian gravity. 
See \cite{DiFandGuit}.} the height model 
we work with contains exactly the same constraints whether we include 
the $\Tr(A^2B^2)$ operator or not. 
These arguments show that this operator is not modifying the essential 
nature of the scalar model living in the continuum. 
Thus the central charge should be the same.
 
Although a complete solution of this matrix model has not been achieved, 
we have been able to explore the renormalization group flow.
We find again the conclusion that the 
gonihedric loop model coupled to gravity corresponds to $c=0$ central 
charge for the matter part\cite{KPZ}, i.e. pure gravity. 
This again agrees with the lack of a
continuous thermodynamical transition observed for the two-dimensional 
gonihedric spin model. We therefore find a rather nice agreement among
all approaches.

\section{Conclusions}

We have presented two, quite different, versions of
the gonihedric loop model.

First, through a direct duality we have transformed our gonihedric spin 
model into a loop model. We have studied the gonihedric loop model 
through the height models using the results given in \cite{DiFandGuit,kondev}. 
This height models has been proved to be give very satisfactory 
predictions for the central charge and critical exponents for 
combinatorial loop models. 

We have applied height model technology to the present problem
and we have extracted the central charge in some limiting cases 
of the gonihedric loop model. These limiting cases correspond to
 zero and infinity temperature, and  zero and infinity $\kappa$ (self-avoidance
 parameter). Modulo some very plausible hypothesis concerning the
interpretation of the central charge and renormalization-group flows,
 we are then able to 
conjecture the value for the central charge found for the limiting models 
to the whole temperature and $\kappa$ space. The value of the central charge 
happens to be the same for the whole space of parameters ($c=0$), 
so we find no continuous transition for the gonihedric model (in agreement 
with existing numerical simulations\cite{esp-pra}).

Secondly, we have coupled our loop model to gravity 
in the $\kappa\to\infty$ limit. We have worked out a partial 
integration since in this limit the {\it loop} matrix is Gaussian. 
Unfortunately we could no go further in the exact 
integration of this matrix model due to its strong non-locality (in 
eigenvalue space).

A renormalization group approach has proved useful to bypass this problem.
We have computed the infinitesimal renormalization flow for the 
two couplings $g$, and $\lambda$. We found that they flow always 
towards zero, so only a quartic one-matrix model remains. 
We can also consider only $\lambda$ as irrelevant coupling constant. 
The limiting matrix model has been already solved and it turns out to 
correspond  to pure gravity.
Then using the  KPZ\cite{KPZ} relations we rediscover the fact that $c=0$ 
for the two-dimensional gonihedric spin model.

\section*{Acknowledgements}
D.E. wishes to thank J. Alfaro and  V. Kazakov for discussions and the 
Department of Physics of the Universidad Catolica de Chile for hospitality.  
A.P. wants to thank F. Di Francesco for interesting discussions in 
SPhT-Saclay and during the short program 
 'Random Matrices, Random Processes and Integrable Systems' in 
Universite de Montreal. 
A.P. acknowledges the support of the CIRIT grant 2001FI-00387. The financial 
support of grants FPA2004-04582, SGR2001-00065 and MRTN-CT-2004-005616 
(European Network ENRAGE) is acknowledged.

\end{document}